# Characterizing Multimedia Information Environment through Multi-modal Clustering of YouTube Videos

Niloofar Yousefi, Mainuddin Shaik, and Nitin Agarwal

COSMOS Research Center, University of Arkansas - Little Rock, AR 72204, USA
{nyousefi,mxshaik,nxagarwal}@ualr.edu

**Abstract.** This study aims to investigate the comprehensive characterization of information content in multimedia (videos), particularly on YouTube. The research presents a multi-method framework for characterizing multimedia content by clustering signals from various modalities, such as audio, video, and text. With a focus on South China Sea videos as a case study, this approach aims to enhance our understanding of online content, especially on YouTube. The dataset includes 160 videos, and our findings offer insights into content themes and patterns within different modalities of a video based on clusters. Text modality analysis revealed topical themes related to geopolitical countries, strategies, and global security, while video and audio modality analysis identified distinct patterns of signals related to diverse sets of videos, including news analysis/reporting, educational content, and interviews. Furthermore, our findings uncover instances of content repurposing within video clusters, which were identified using the barcode technique and audio similarity assessments. These findings indicate potential content amplification techniques. In conclusion, this study uniquely enhances our current understanding of multimedia content information based on modality clustering techniques.

## 1 Introduction

In today's digital world, we are witnessing an extraordinary surge in the production of content in various forms, each potentially impacting users in distinct ways [1, 2]. From social media platforms and streaming services to news outlets and educational institutions, video has established itself as the dominant medium for communication, entertainment, and information dissemination [3]. The reasons for this are manifold: videos can convey complex ideas in an engaging manner, capture the nuances of human emotion, and bring stories to life in a way that text or images alone cannot [4, 5].

With the proliferation of smartphones, tablets, and affordable recording devices, creating and sharing videos has never been easier. Platforms like YouTube,

TikTok, Instagram, and Twitter have played pivotal roles in catalyzing this explosion of video content, leading to a scenario where, every minute, hundreds of hours of video are uploaded to the internet [6, 7].

In this avalanche of video data, a crucial challenge arises: How can we effectively manage and make sense of this vast repository of information? This is where the importance of multimedia information characterization comes into play. It pertains to the techniques and methodologies employed to extract, analyze, and categorize video data. Whether it's for the purpose of video recommendation, content moderation, or anomaly detection in video streams, understanding the content of a video is fundamental to ensuring its accessibility, relevance, and usefulness to viewers [8]. Additionally, in the midst of this flood of video content, another major issue arises as many creators are repurposing content on platforms like YouTube. Detecting these instances is vital for preserving authenticity and credibility online [9].

Our research enhances the existing body of literature by developing a multi-method framework to characterize the information content from videos, particularly those on YouTube. We achieve this by clustering signals from different modalities, such as audio (extracting information from the video's soundtrack), visual (extracting video objects, scenes, and colors), and text (extracting and understanding any on-screen text or captions). This study builds upon the concept of multimedia information characterization to gain insights and detect content cloning or repurposing through an efficient video analysis framework.

A significant portion of research on multimedia information characterization has mainly focused on modalities such as text [10–12] and visual [15, 14, 13] with few studies emphasizing audio [17, 18, 16]. However, no study to date has combined all these modalities to effectively characterize the information from YouTube video comprehensively. In the next section, we review prior research in the realm of video information characterization. This review aids in explaining our novel framework, offering a better understanding of video analysis.

## 2   Literature Review

Video Information Characterization (VIC) can be defined as the process through which pertinent information is extracted from the video content, analyzed, and subsequently categorized or tagged based on its content, context, or semantics. The scope of video information characterization is vast, given the multimedia nature of YouTube video content [8]. It spans from basic methods of video tagging based on metadata to more complex analyses such as visual, auditory, and anomaly detection in video streams.

**Visual characterization:** *Non-YouTube:* One study introduced the Video-Book framework for content-based video retrieval. This framework characterizes video sequences using an 8-dimensional vector derived from motion, texture, and colorimetry. Impressively, it achieves a 92% accuracy rate in real-time retrieval. The method draws inspiration from the human visual cortex and employs mutual information for feature extraction [19]. Concurrently, another research focused on understanding the higher-level structure in videos to achieve meaningful characterization. This research introduced a Bayesian approach to determine shot boundaries and adopted a two-state model for videos. It also underscores the significance of shot duration and activity in video structure, hinting at a broader semantic characterization in future studies [14]. A similar study introduced "pictorial summaries" as a novel approach to visually representing videos. These summaries comprise representative images or "video posters" that capture the video's unique elements. The paper defines "story units" in videos and explores the synergy between video characterization and compaction. It emphasizes the potential of pictorial summaries in the field of digital libraries [20].

*YouTube:* A couple of studies have delved into the characterization of YouTube videos from a traffic-on-network infrastructure perspective. By using a customized web spider, one research collected metadata from popular YouTube videos. It underscored the significance of understanding video metadata and viewing history for traffic optimization strategies [21]. Another research investigated the characteristics of popular YouTube videos, analyzing the top most viewed videos across various time frames. This research investigates video characteristics, popularity trends, and category preferences. It offers insights into user behavior and content popularity [15]. More recent advancements in the field of visual characterization on YouTube suggested barcode technique that uses color theory to summarize videos by compressing an entire video into one image. This technique identifies similarities among videos and categorizes them [22].

**Audio characterization:** *Non-YouTube:* For audio characterization from video, one study investigates the MPEG encoding process and introduces methods to differentiate between silent and non-silent segments. It showcases a pitch detection algorithm and evaluates their approach using a movie segment [17].

*YouTube:* This study explored the discovery of recurring audio events in a collection of YouTube videos. Using a streaming, non-parametric clustering algorithm, the authors demonstrate the effectiveness of the discovered audio event clusters in weakly-supervised learning and informative activity detection [18].

**Textual characterization:** *Non-YouTube:* A Couple of generic studies focused on clustering videos based on transcripts used Latent Dirichlet Allocation (LDA) techniques for different purposes. One study used academic video content. It uses the transcripts of lecture videos to represent the content and employs LDA for topic modeling. The Kullback-Leibler divergence (KL Div) is used to measure the similarity between topic distributions, and Fuzzy C-Means clustering is applied for the clustering of videos [10]. Another research presents a method for indexing and retrieving video transcripts. The data analysis pipeline involves transforming transcripts into a Bag of Words format, computing TF-IDF scores, clustering using k-means, and building an LDA model for topic identification. The paper emphasizes the potential of this method for efficient retrieval of relevant video transcripts, especially in the context of Massive Open Online Courses and Technology-Enhanced Learning (TEL) systems [23].

*YouTube:* Few studies looked into different types of textual data from YouTube by performing topic modeling techniques to characterize them with a focus on enhancing the user experience on online platforms. One example is to identify abusive comments on YouTube videos. In this study, authors investigate toxic behavior in YouTube comments. Using the YouTube data API v3 and LDA, the study identifies dominant topics in comments and employs sentiment analysis [11]. Another study introduced "Videopedia," a system designed to enhance e-learning blogs by recommending relevant lecture videos. The system uses LDA topic modeling, on video transcripts to determine the similarity between blogs and videos, subsequently recommending pertinent videos [24].

**Multiple modality characterization:** *Non-Youtube:* One study delved into the multimedia content analysis of music video clips, aiming to extract effective information. This research emphasizes the emotional impact of multimedia content, combining features from audio and video streams. It uses a multilevel fusion scheme to capture emotions [25]. Another research underscored the importance of integrating audio and video information for efficient video indexing. In this study, the authors proposed a broader definition of a "scene" and segmented audio into various classes. They noted that audio often precedes video changes, suggesting that a combined audio-video analysis can enhance scene detection [26]. Similarly, another study unveiled a video skimming technique designed to offer an overview of the original video content. This method merges audio and image characterization, incorporating face and text detection, to extract pivotal information and craft a "skim" video [27].

In a review of the existing literature, it is evident that prior studies have employed diverse methodologies and frameworks to extract information from multimedia, be it from YouTube or other platforms. Notably, no study to date has integrated all modalities—visual, audio, and text—within a single research endeavor to holistically cluster signals from these varied modalities, thereby enhancing the characterization of information from videos. Our primary contribution to the field lies in addressing this gap, as we aim to enhance the characterization of information from videos by integrating multiple modalities. Additionally, our secondary contribution involves the introduction of a novel framework that incorporates a multi-method analysis, facilitating the identification and detection of content cloning or repurposing. In sum, our research not only augments the current body of knowledge on information characterization from multimedia but also unveils new pathways for delving into multimedia content. This naturally segues into our research questions. **RQ1:** How do different modalities help in characterizing the information? **RQ2:** How does multi-modal analysis enhance the detection of content repurposing tactics in online video content?

## 3   Methodology

In this section, we outline our methodology for text, video, and audio assessments, along with the data collection process.

### 3.1 Data Collection

For this study, we collected data by utilizing YouTube API keys and employing a set of keywords related to the South China Sea [30]. Some of the finalized set of keywords are: "West Philippine Sea"/"South China Sea", "South China Sea"/SCS/"West Philippine Sea"/WPS + China/Philippine + militia

The data was collected using the specified keyphrases provided by subject matter experts to ensure their high relevance to the subject [31]. A collection of 40 seed videos on YouTube was collected. To expand our dataset, we looked at the related recommended videos. To achieve this, we employed a custom crawler to retrieve five depersonalized video recommendations for each initial seed video, forming the first depth of recommendations. This process was iterated four more times, with videos from the previous depths serving as inputs to generate subsequent depths of recommendations. Depersonalization was a step taken to assess the raw recommendation algorithm's performance without the influence of user-specific history [32]. By employing this methodology, we effectively selected a dataset consisting of 160 related YouTube videos.

### 3.2 Text Extraction

Initially, the study involved using YouTube's Transcript API to extract transcripts generated by YouTube. The OpenAI Whisper model was employed to generate transcriptions for videos that do not have native YouTube-generated transcripts. The transcript collection process from YouTube was optimized with parallel computing and the multiprocessing library. Additionally, Googletrans Translation API was used to translate non-English video transcriptions [33].

Next, the BERT model is employed to generate text embeddings, representing input text as numerical vectors. Subsequently, Text similarity was computed using cosine similarity on these embeddings, and K-means clustering was applied to group similar texts. K-means groups textual data into clusters, ensuring that the content similarity among texts within a cluster surpasses that of the texts in different clusters. K-means is widely acknowledged as an effective text clustering algorithm, and it was chosen for the clustering task [28, 29]. The input comprises the video title, transcript, and description. To identify key topics within the clusters, we used Latent Dirichlet Allocation (LDA) for topic modeling. We initiated the process by breaking down tweets into sentences, further tokenizing sentences into words while removing punctuation and common stop-words. Initially, the LDA model underwent training with various topic counts, which were subsequently narrowed down to the top 3 topics based on their coherence scores.

### 3.3 Video Barcode Extraction

Video barcode is a technique designed to transform videos into unique images. Videos comprise of a series of frames, typically spanning 30 to 60 frames per second. The core idea of the video barcode concept is the extraction of dominant colors from each frame. This process begins by isolating individual frames from the video. For each frame, the average values of the Red (R), Green(G), and Blue(B) color channels are calculated, resulting in a three-value vector (RGB). These vectors are then stacked together to form an RGB matrix, effectively representing the entire video. Subsequently, the RGB matrix is converted into a tensor, so it can be visualized as an image. The width of this image corresponds to the number of frames in the video, while the length is typically set to a user-defined value, 224 pixels in our experiments. Figure 1 indicates a sample barcode, each barcode comprises the dominant colors from each frame. The sequence of colors and their changes over time can convey important information about the video, such as the main focus, the subject, and the video's narratives [22]. Each video has its unique barcode. This means that if the same scene is captured with the same camera from different angles, the resulting barcodes will be different.

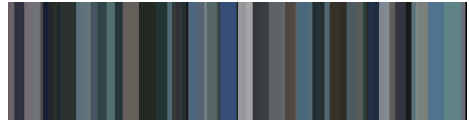

Fig. 1: A sample video barcode image of China news.

### 3.4 Audio Feature Extraction

The audio assessment describes the process of extracting information from audio signals for analysis. Feature extraction is employed to transform unstructured audio compositions into organized data inputs. Librosa is a Python library for analyzing audio files. It provides a set of tools to work with audio data and features such as loading audio data from various file formats, including WAV. It returns a tuple containing the audio data as an array and the sampling rate of the audio. The sampling rate is the number of data points taken per second. Next, extract the Mel-frequency cepstral coefficients (MFCC) feature which is a representation of the short-term power spectrum of an audio signal. It captures important characteristics of the audio, such as the shape of the spectral envelope, which is related to the way humans perceive sound. MFCCs are often used for tasks like speech recognition, and audio classification. They are particularly useful for capturing the timbral (tonal and textural) characteristics of audio, making them suitable for tasks where understanding the content of the audio is essential. Another Librosa tool provides a set of functions for visualizing audio data, such as waveforms. A waveform is a representation of a signal that shows how the signal's amplitude varies over time. The amplitude of a wave is the maximum deviation of a waveform from its centerline to its peak, which determines the level of pressure variation in its surroundings. The greater the amplitude, the higher the pressure variation and, therefore the louder the sound perceived.

## 4 Results and Findings

In this section, we discuss the findings and results of the clustering analysis based on audio, video barcode, and text data. Our content analysis in these three modalities clustering will address our **research question one**.

### 4.1 Text-based Clustering

For text-based similarity analysis, each video was associated with a composite feature comprising the video's title, description, and transcript, which were extracted using the YouTube API. Subsequently, the similarity of this composite feature was computed, as explained in Section 3.2. After obtaining the results of the similarity measurements, we determined the optimal number of clusters, K, for our K-means clustering process using both the Silhouette score and the Elbow method. The most suitable number of clusters was found to be two. We then performed topic modeling on each of these two clusters to gain a deeper understanding of the thematic focus within each cluster. Within this context, we extracted the top three topics for each cluster, with each topic comprising a list of ten associated words Based on the results of the topic modeling, it appears that cluster 1 is likely related to the following topics:

1. Taiwan and Philippine Islands: It mentions "Taiwan," "Philippines," and "island." It also references "navy," which might suggest discussions related to maritime security or naval activities in the South China Sea.
2. Geopolitical Concerns and Strategies: It appears to revolve around geopolitical concerns, strategies, and international relationships. It mentions "Russia" and terms like "strategic". This topic may be about the involvement of countries like Russia in the South China Sea issue.
3. Global Policy and Security: It mentions "global," "policy," and "security." It's likely that this topic deals with discussions about global responses to issues in the South China Sea.

This cluster appears to be touching on a range of topics related to the South China Sea, including specific countries and islands, geopolitical strategies and concerns, and global policy and security considerations.

In the second cluster, overall, the main topic extracted is mostly centered around international geopolitics. with a focus on issues such as war, security, government, and the involvement of countries like India, Ukraine, the United Kingdom, and Russia. Notably, posts in this cluster display less relation with the South China Sea issue compared to the first cluster. Our data collection process, as previously outlined, began with a basic set of initial videos, which served as the starting point for recommendations, referred to as 'depth 0'. This process generates four subsequent recommendation depths. It's noteworthy that at depth 0, there is a significant association with South China Sea-related content. However, as we delve deeper into subsequent recommendation levels, we observe a gradual shift in focus toward international geopolitics. We observe shifts in themes as we explore deeper recommendations [34].

Through our textual analysis-based clustering approach, we observe the first cluster mostly includes content from initial depths, Highlighting a focus on the South China Sea. In contrast, the second cluster is primarily populated with content from subsequent depths, wherein the thematic focus shifts to international geopolitics, highlighting a transition in content themes at these deeper recommendation depths.

### 4.2   Barcode-based Clustering

For this study, barcodes were generated for the videos, employing the methodology explained in Section 3.3. The barcode images were processed, leading to the conversion of image data in the form of numpy arrays. Subsequently, employing silhouette scores and the Elbow method, the optimal number of clusters for K-means clustering was determined to be three. The K-means clustering algorithm was then applied to the entire dataset, This led to associating video IDs to their respective clusters. In addition, the RGB values were extracted for each video ID in each cluster. Figure 2 illustrates the color conversion based on the average RGB values of the video barcodes within each cluster. In the following step, for the purpose of cluster profiling, a thorough examination of each cluster was performed. Cluster profiling involves thoroughly analyzing the unique characteristics and attributes of each cluster, which helps identify distinct patterns and features within the dataset.

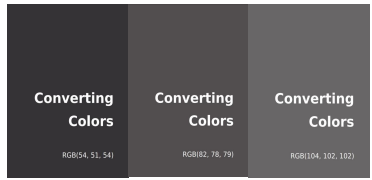

Fig. 2: Color conversion based on the average RGB values for each cluster.

The main focus within the **first cluster** (distinguished by its lighter average color) is mainly centered around educational and information-centric content. The content within this cluster comprised informative presentations, explanatory materials, and educational resources designed to increase understanding and knowledge dissemination regarding the South China Sea. Figure 3a indicates some examples of the videos that were assigned to this cluster.

The videos within **second cluster** predominantly feature interview dialogues among individuals, news channel reporters engaging in discussions, or single individuals providing explanatory narratives. In figure 3b some examples of such footage are displayed.

The **third cluster** primarily comprises videos depicting events, or news segments where reporters provide commentary while showcasing footage of urban environments. This cluster underscores the prevalence of dynamic and on-site reporting, capturing real-time occurrences and cityscapes. Figure 3c presents some examples of videos from the third cluster.

Subsequently, to answer our **research question two,** we undertook a more advanced analysis by examining a set of samples, in which we encountered videos displaying similarities. Our investigation revealed instances of content repurposing within these videos. Content repurposing is a tactic where an individual or entity, like a YouTube channel, takes content, such as a video, either partially or entirely, from other sources. The similarity between the content can vary, ranging

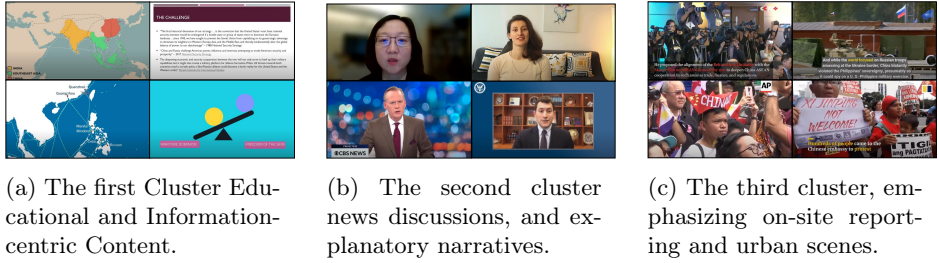

(a) The first Cluster Educational and Information-centric Content.

(b) The second cluster news discussions, and explanatory narratives.

(c) The third cluster, emphasizing on-site reporting and urban scenes.

Fig. 3: The characteristic of different clusters.

from highly similar to completely identical. Using our video barcode methodology to identify content similarity, as depicted in Figure 4, wherein the barcode of a video pair reveals segments of high similarity. In Figure 5, we provide a visual representation of a pair of videos that exhibit a high degree of similarity. Using barcode-based video similarity assessment can be a significant tool for identifying clusters of videos that reuse or repurpose content from other videos. This practice of content cloning or repurposing offers two key outcomes. First, it expands the content's audience, enhancing the probability of it being suggested by the platform's content recommendation and related feed curation systems. Second, in situations where the platform takes down or suspends offensive content due to policy violations, having a duplicate of the content guarantees continued availability and distribution.

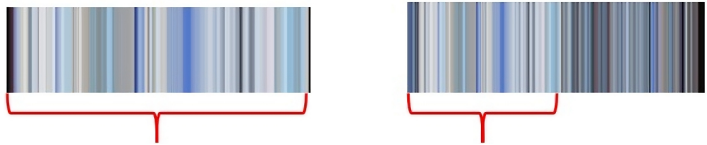

Fig. 4: Barcode of a video pair reveals segments of high similarity.

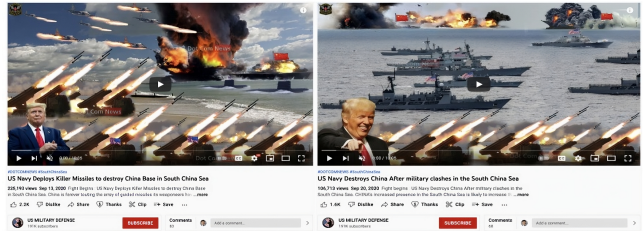

Fig. 5: Picture frame of a pair of videos with high similarity.

### 4.3  Audio-based Clustering

In the process of audio assessment, we initiated by extracting audio features using Librosa. Specifically, we extracted an array containing the raw audio data along with MFCC features. Subsequently, we calculated the cosine similarity among the audio files and applied the K-means clustering algorithm to cluster the audio data. However, prior to this clustering step, we determined the optimal number of clusters for K-means through the application of the Elbow method and Silhouette score analysis. Our analysis revealed that the most suitable number of clusters for this task was two, and as such, the K-means algorithm segregated the audio dataset into two distinct clusters.

Within **first cluster**, a notable observation was the videos characterized as podcasts or news with significant usage of microphones or instances where a speaker engaged in verbal discourse within a video depicting an event, accompanied by background music. Additionally, there were videos with notably loud introductions. These various audio attributes and characteristics were found to significantly influence the waveform patterns within this cluster. The **second cluster** primarily encompasses videos characterized by ordinary conversations, interviews between individuals, and instances in which a single individual engages in verbal communication with an audience, whether it be in a public forum, or within a professional context. All of these are often characterized by audio quality that may not meet high standards. Figure 6 and 7 indicate instances of waveforms within each cluster. The Y-axis represents the amplitude of the audio signal, which ranges from -1 to 1. The X-axis represents the time.

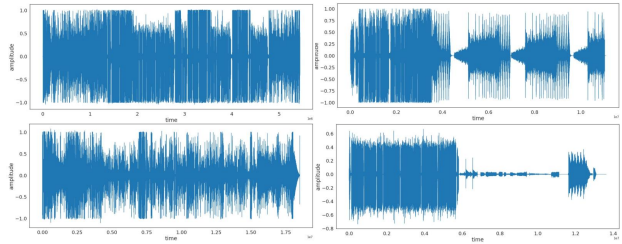

Fig. 6: Instances of waveforms originating from the first cluster.

Next, we conduct a more in-depth analysis of the videos within the identified clusters that exhibit similarities. By analyzing the waveforms, we discovered that some videos incorporate similar or identical audio segments from other sources, which will address our **research question two**. In Figure 8-a the waveforms show the presence of identical segments in the waveforms of the two signals. Manual checking confirmed that both videos use identical introductions and closing remarks. Figure 8-b, shows that there are two similar videos from different channels on YouTube, these videos share the same audio but with slight differences. The analysis of the second video suggests that it uses the same footage as the first video. This observation is significant, as it suggests that the second video may have been created by copying or reusing content from the first video without proper attribution or authorization.

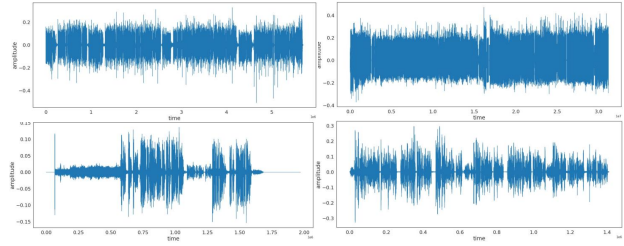

Fig. 7: Instances of waveforms originating from the second cluster.

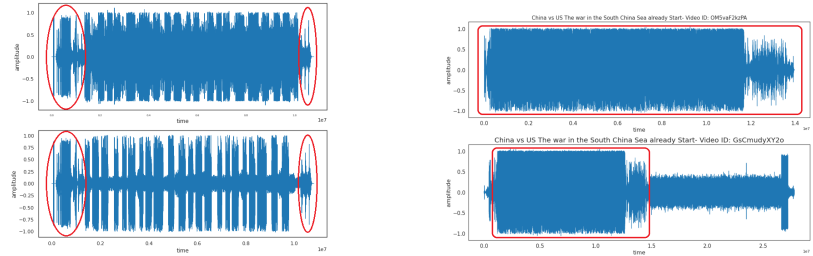

(a) Identical segment in two waveforms.   (b) Using Identical video in another video.

Fig. 8: Presence of identical segments in the waveforms of the two signals.

## 5  Conclusions and Future Works

This study investigates a comprehensive analysis of videos, employing a multi-method approach based on audio, video, and text. The findings from this analysis interpret the diverse content and themes, offering valuable insights into the content and providing a more comprehensive perspective that single-modal analyses might miss. This research presents two significant contributions to the field of multimedia information characterization:

**Multimodal Content Analysis:** Our clustering approach harnesses multiple modalities to delve into content. Text-based clustering elucidates thematic paradigms, highlighting predominant themes in South China Sea content: regional countries, geopolitical maneuvers, and global policy-security considerations - invaluable to policymakers and South China Sea enthusiasts. Meanwhile, barcode-based clustering furnishes a novel method for visual analysis, segmenting video content based on color characteristics, aiding content identification and categorization. This method discerns three distinct clusters: educational content, interviews and explanatory segments, and on-site reporting. Furthermore, audio-based clustering, hinging on audio features, surfaces two distinct clusters, distinguished by audio quality and amplitude. Such a multifaceted approach offers a comprehensive understanding of the multifarious themes and topics tied to the South China Sea.

**Detection of Content Repurposing:** Our meticulous analysis unveils instances of content repurposing within video clusters. The video barcode methodology discerns segments with striking similarities, while our audio-based scrutiny identifies videos with shared content and pinpoints identical segments across varied channels. These findings not only spotlight potential content amplification techniques and unauthorized content duplication on social media platforms but also underscore the implications for content recommendation, intellectual property rights, and platform policy enforcement.

Future work can apply this methodology to larger datasets and alternative clustering methods and also apply graph neural networks [35]. However, Collecting more data may pose challenges, as manual validation of associated keywords for data collection can be time-intensive. Furthermore, we will explore leveraging learning techniques to improve and optimize performance, thereby making it easier to apply our algorithms to larger datasets. Additionally, merging our clustering techniques into a unified model for comprehensive video categorization, along with combining multimodal to create one feature vector representation, will enhance our work. is also envisaged, along with analyzing user engagement to find out what content and themes are most popular.

## Acknowledgement

This research is funded in part by the U.S. National Science Foundation (OIA-1946391, OIA-1920920, IIS-1636933, ACI-1429160, and IIS-1110868), U.S. Office of the Under Secretary of Defense for Research and Engineering (FA9550-22-1-0332), U.S. Office of Naval Research (N00014-10-1-0091, N00014-14-1-0489, N00014-15-P-1187, N00014-16-1-2016, N00014-16-1-2412, N00014-17-1-2675, N00014-17-1-2605, N68335-19-C-0359, N00014-19-1-2336, N68335-20-C-0540, N00014-21-1-2121, N00014-21-1-2765, N00014-22-1-2318), U.S. Air Force Research Laboratory, U.S. Army Research Office (W911NF-20-1-0262, W911NF-16-1-0189, W911NF-23-1-0011, W911NF-24-1-0078), U.S. Defense Advanced Research Projects Agency (W31P4Q-17-C-0059), Arkansas Research Alliance, the Jerry L. Maulden/Entergy Endowment at the University of Arkansas at Little Rock, and the Australian Department of Defense Strategic Policy Grants Program (SPGP) (award number: 2020-106-094). Any opinions, findings, and conclusions or recommendations expressed in this material are those of the authors and do not necessarily reflect the views of the funding organizations. The researchers gratefully acknowledge the support.